\documentclass[aps,prb,twocolumn,superscriptaddress,notitlepage]{revtex4-1}
\usepackage{graphicx}
\usepackage[caption=false]{subfig}
\usepackage{float}
\usepackage{natbib}
\usepackage{xcolor}
\usepackage{xr}
\usepackage{amsmath}
\usepackage{amssymb}
\usepackage{array}
\usepackage{wasysym}
\usepackage{color,soul}
\usepackage{braket}
\usepackage{verbatim}
\usepackage{hyperref}
\usepackage{gensymb}
\usepackage{url}
\usepackage{dirtytalk}

\usepackage{filecontents}
\begin{document}

\title{Tailoring the normal and superconducting state properties of ternary scandium tellurides, Sc$_6M$Te$_2$ ($M = $ Fe, Ru and Ir) through chemical substitution}


\author{J. N. Graham}
\email{jennifer.graham@psi.ch}
\affiliation{PSI Center for Neutron and Muon Sciences CNM, 5232 Villigen PSI, Switzerland}

\author{K. Yuchi}
\affiliation{Institute for Solid State Physics, University of Tokyo, Kashiwa, Chiba 277-8581, Japan}

\author{V. Sazgari}
\affiliation{PSI Center for Neutron and Muon Sciences CNM, 5232 Villigen PSI, Switzerland}

\author{A. Doll}
\affiliation{PSI Center for Neutron and Muon Sciences CNM, 5232 Villigen PSI, Switzerland}

\author{C. Mielke III}
\affiliation{PSI Center for Neutron and Muon Sciences CNM, 5232 Villigen PSI, Switzerland}
\affiliation{Microstructured Quantum Matter Department, Max Planck Institute for the Structure and Dynamics of Materials, Luruper Chaussee 149, 22761 Hamburg, Germany}

\author{P. Král}
\affiliation{PSI Center for Neutron and Muon Sciences CNM, 5232 Villigen PSI, Switzerland}

\author{O. Gerguri}
\affiliation{PSI Center for Neutron and Muon Sciences CNM, 5232 Villigen PSI, Switzerland}

\author{S.S. Islam}
\affiliation{PSI Center for Neutron and Muon Sciences CNM, 5232 Villigen PSI, Switzerland}

\author{V. Pomjakushin}
\affiliation{PSI Center for Neutron and Muon Sciences CNM, 5232 Villigen PSI, Switzerland}

\author{M. Medarde}
\affiliation{PSI Center for Neutron and Muon Sciences CNM, 5232 Villigen PSI, Switzerland}

\author{H. Luetkens}
\affiliation{PSI Center for Neutron and Muon Sciences CNM, 5232 Villigen PSI, Switzerland}

\author{Y. Okamoto}
\affiliation{Institute for Solid State Physics, University of Tokyo, Kashiwa, Chiba 277-8581, Japan}

\author{Z. Guguchia}
\email{zurab.guguchia@psi.ch} 
\affiliation{PSI Center for Neutron and Muon Sciences CNM, 5232 Villigen PSI, Switzerland}

\date{\today}

\begin{abstract}
The pursuit of a unifying theory for non-BCS superconductivity has faced significant challenges. One approach to overcome such challenges is to perform systematic investigations into superconductors containing \textit{d}-electron metals in order to elucidate their underlying mechanisms. Recently, the Sc$_6M$Te$_2$ ($M$ = \textit{d}-electron metal) family has emerged as a unique series of isostructural compounds exhibiting superconductivity across a range of $3d$, $4d$, and $5d$ electron systems. In this study, we employ muon spin rotation, neutron diffraction, and magnetisation techniques to probe the normal and superconducting states at a microscopic level. Our findings reveal extremely dilute superfluid densities that correlate with the critical temperature ($T_\mathrm{c}$). Additionally, we identify high-temperature normal-state transitions that are inversely correlated with $T_\mathrm{c}$. Notably, in Sc$_6$FeTe$_2$, the superconducting pairing symmetry is most likely characterised by two nodeless gaps, one of which closes as electron correlations diminish in the Ru and Ir Sc$_6M$Te$_2$ compounds. These results classify the Sc$_6M$Te$_2$ compounds ($M$ = Fe, Ru, Ir) as unconventional bulk superconductors, where the normal-state transitions and superconducting properties are governed by the interplay between electron correlations and spin-orbit coupling of the \textit{d}-electron metal.


\end{abstract}
\maketitle
\section{Introduction}
There is a great diversity of $d$-electron metals across the periodic table whose physical properties are largely dictated by the strength of their electron correlations. In some instances, these correlations can lead to superconducting states with examples including the high-temperature cuprates \cite{doi:10.1126/science.237.4819.1133,keimer2015quantum,Uemura1, greene2020strange, rahman2015review, hayden2023charge}, strontium ruthenate Sr$_2$RuO$_4$ \cite{10.1063/1.1349611,mackenzie2003superconductivity,grinenko2021split}, kagome $A$V$_3$Sb$_5$ ($A = $ K, Rb, Cs) \cite{AV3Sb5_1, AV3Sb5_chiral3,kagome_vH3,mielke2022time,guguchia2023unconventional,graham2024depth} compounds and the recently discovered nickelates \cite{wang2024experimental, nomura2022superconductivity}. Despite these numerous examples it remains unpredictable as to which specific combination of $d$-electron metal and chemical structure will generate a superconducting ground state. Subsequently, the inability to perform systematic studies of $d$-electron substitution within a single material family has likely prevented the formation of a unifying theory to the generic aspects behind $d$-electron based superconductivity.

Recently, a new family of superconductors were identified, the ternary scandium tellurides, Sc$_6M$Te$_2$, in which species from all $3d$, $4d$ and $5d$ electron systems could be incorporated into the $M$ site \cite{maggard2000sc6mte2, chen2002synthesis, Sc6MTe2}.  The Sc$_6M$Te$_2$ compounds adopt the hexagonal $P\bar{6}2m$ structure without inversion symmetry which is comprised of $M$ atoms co-ordinated by a distorted trigonal prismatic environment of six Sc atoms. These Sc$_6M$ clusters are then connected in one-dimensional chains along the $c$-axis (Fig. 1a). At interstitial positions, Te atoms form a layered hexagonal net around the Sc$_6M$ chains (Fig. 1b). 
\begin{figure}
    \centering
    \includegraphics[width=0.5\textwidth]{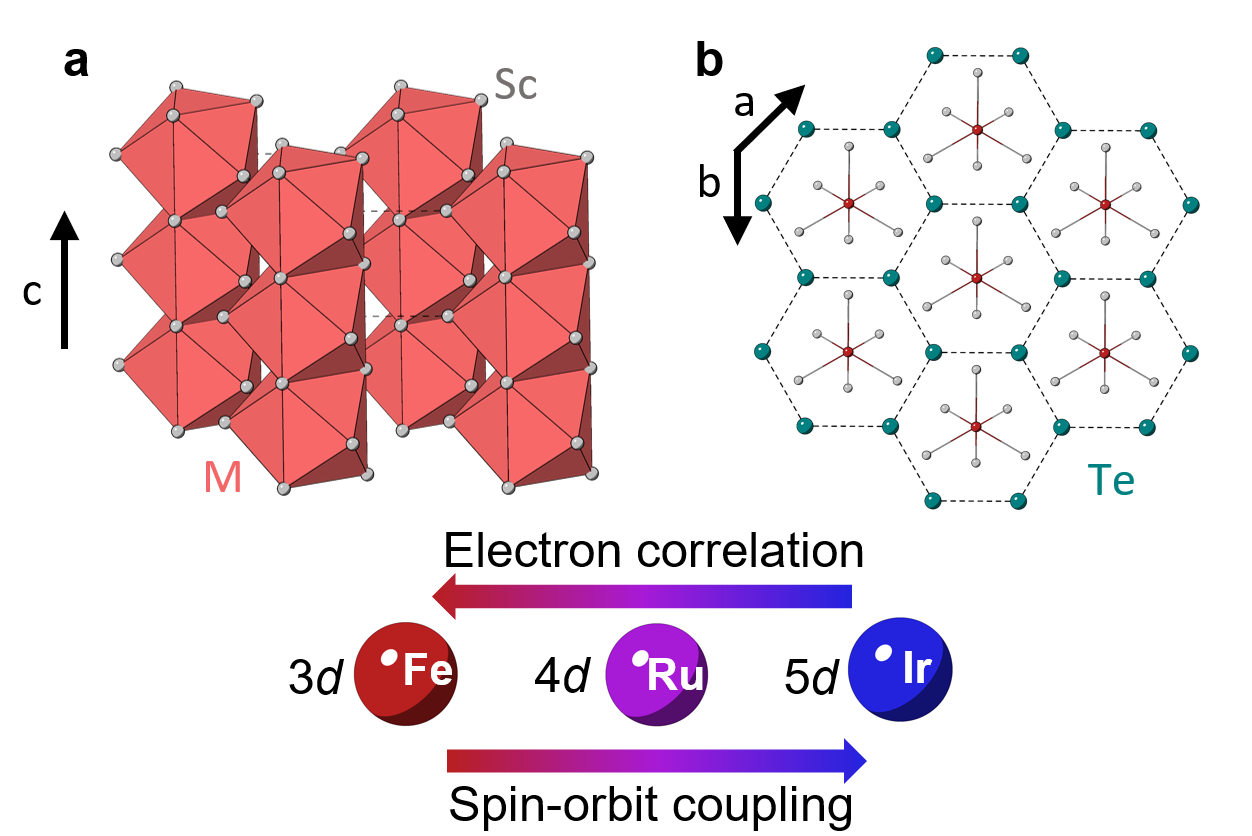}
    \caption{\textbf{Crystal structure of Sc$_6M$Te$_2$ ($M = $ Fe, Ru and Ir). a} The Sc$_6M$Te$_2$ materials adopt the $P\bar{6}2m$ spacegroup without inversion symmetry where the $M$ atoms are co-ordinated by a distorted trigonal prismatic environment that form a 1D chains along the $c$-axis. The flexibility of the structure allows for the substitution of a number of $d$-electron metals, specifically in this study, $3d$-Fe, $4d$-Ru and $5d$-Ir, which were chosen for their varying degrees of spin-orbit coupling and electron correlations. \textbf{b} In the $ab$ plane, Te atoms form a hexagonal sublattice around the $Sc_6M$ chains.}
    \label{fig1}
\end{figure}
Chemical substitution into the $M$ site does not change the chemical structure and, remarkably, examples of superconductivity were found across the entire $3d$, $4d$, and $5d$ series \cite{Sc6MTe2}. This is a highly unusual and novel characteristic for a $d$-electron based family, and therefore the Sc$_6M$Te$_2$ compounds are an exciting new addition to the pool of superconducting materials. 

\begin{figure*}
    \centering
    \includegraphics[]{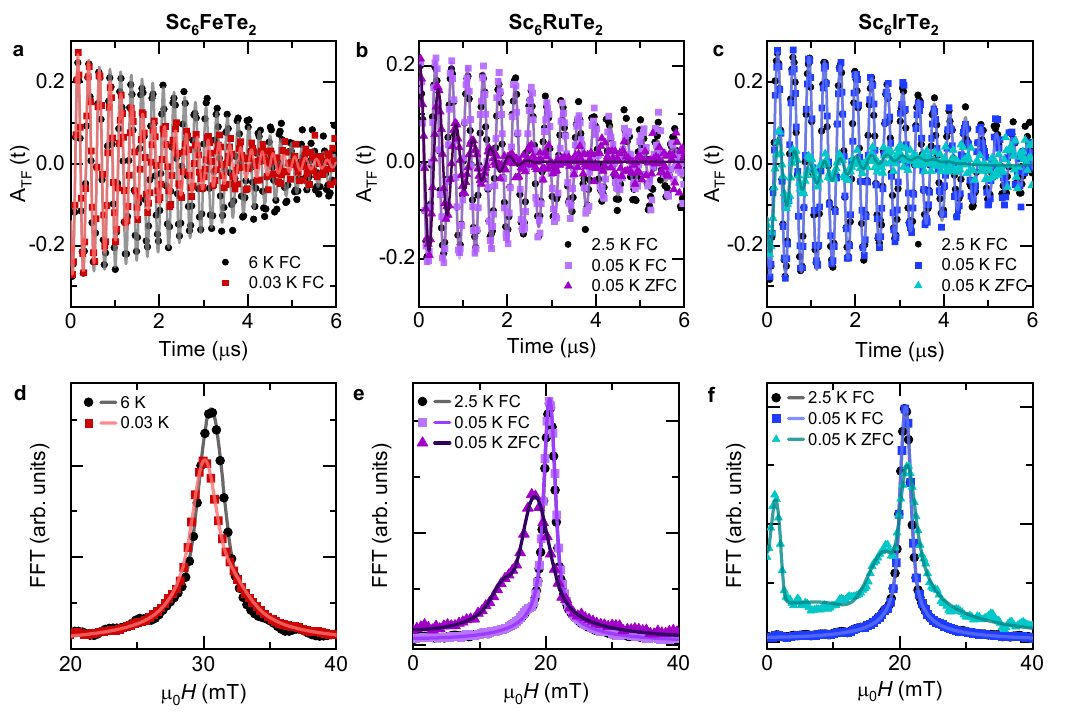}
    \caption{\textbf{Transverse-field (TF) $\mu$SR time spectra and their corresponding Fourier transforms for Sc$_6M$Te$_2$ ($M = $ Fe, Ru and Ir).} \textbf{a - c} TF-$\mu$SR spectra collected above (black) and below (coloured) $T_\mathrm{c}$ after field-cooling the sample from above $T_\mathrm{c}$. For the $M = $ Ru and Ir samples, an additional measurement is included after a zero-field cooling cycle. Measurements were conducted in a $30~$mT applied field for Sc$_6$FeTe$_2$ and $20~$mT applied field for Sc$_6$RuTe$_2$ and Sc$_6$IrTe$_2$. \textbf{d - f} Real part of Fourier transforms of $\mu$SR spectra. }
    \label{Fig2}
\end{figure*}

The critical temperatures, $T_\mathrm{c}$ were determined by resistivity, magnetic susceptibility and heat capacity measurements to be approximately $2~$K for the $M = 4d$ and $5d$ systems such as Ru and Ir, but higher, with systematic variation according to their atomic number, for $M = 3d$ elements like Fe ($T_\mathrm{c} = 4.7~$K) \cite{Sc6MTe2}. Furthermore, the upper critical field ($H_{c2}$) was also dependent on the $M$ substitution, but in comparison to $T_\mathrm{c}$, $H_{c2}$ is higher for the $M = 5d$ elements than the $M = 3d$ elements. The upper critical fields for $M = $ Fe ($3d$), Ru ($4d)$ and Ir ($5d$) are $\mu_0H_{c2}(0) = 8.68~$T, $3.55~$T and $5.00~$T, respectively. One possible explanation for this behaviour was given by first principle calculations which suggested that $T_\mathrm{c}$ was dependent on which orbitals dominantly contribute to the electronic states near the Fermi level \cite{Sc6MTe2}. For example, superconductivity appears at $2~$K when the dominant contribution is from the $3d$ Sc orbitals, but for systems with higher critical temperatures there is additional overlap from the $M$ orbitals, such as in the $3d$ Fe case.

These initial studies have shown the Sc$_6M$Te$_2$ compounds to be a rare example of a family of isostructural superconductors with the capability to tailor the superconducting properties through chemical substitution of $d$-electron metals. Therefore, this provides an ideal platform to understand the mechanisms behind $d$-electron superconductivity. To explore the superconducting and the normal state properties on the microscopic level, we have chosen to focus on three compounds, $M = $ Fe ($3d$), Ru ($4d)$ and Ir ($5d$) as these have varying strengths of electron correlations and spin-orbit couplings (Fig. 1). Specifically, spin-orbit coupling increases from $3d$ to $5d$ electron systems, while the strength of electron correlations decreases. Therefore by studying these three compounds we anticipate that we will be able to draw conclusions which will be applicable across the rest of the series. Our study combines muon-spin rotation/relaxation ($\mu$SR), neutron diffraction and magnetic susceptibility to systematically determine how electronic correlations play a role in $d$-electron superconductivity. The results classify the Sc$_6M$Te$_2$ compounds as unconventional bulk superconductors, which each have a competing high-temperature state ($T^*$) that is anticorrelated to the superconductivity. We discuss the intriguing consequences of such competition, and how this is related to the substitution of different $d$-electron metals. By advancing this discussion, we hope this may lead to a deeper understanding of the generic aspects behind non-BCS behaviour which is observed in unconventional superconductors, and how to tailor superconducting properties through chemical substitution for future material design.  

\section{Unconventional superconducting state}
\begin{figure*}
    \centering
    \includegraphics[width=\textwidth]{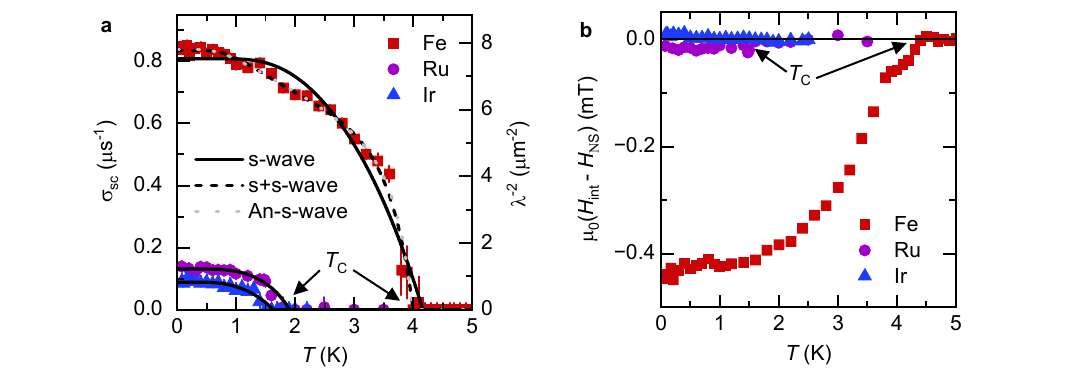}
    \caption{\textbf{Summary of superconducting properties for the Sc$_6M$Te$_2$ ($M =$ Fe, Ru, Ir) compounds. a} Temperature dependence of the superconducting muon spin depolarisation rate, $\sigma_\mathrm{sc}$ (left axis) and inverse squared penetration depth, $\lambda^{-2}$ (right axis) measured in $30~$mT and $20~$mT applied field for the Fe and Ru/Ir samples, respectively. \textbf{b} Response of the internal magnetic field, $\mu_0H_\mathrm{int}$ in the superconducting state. Results are shown in terms of the difference from the applied magnetic field, $\mu_0H_{\mathrm{NS}}$. The Fe and Ru Sc$_6M$Te$_2$ samples have the expected diamagnetic responses, but Sc$_6$IrTe$_2$ has a slight paramagnetic shift below $T_\mathrm{c}$. }
    \label{fig3}
\end{figure*}

We explored the microscopic superconducting properties of the Sc$_6M$Te$_2$ ($M = $ Fe, Ru and Ir) compounds with transverse-field (TF) $\mu$SR measurements which are summarised in Figs. \ref{Fig2} and \ref{fig3}. TF-$\mu$SR is a powerful experimental tool to measure the magnetic penetration depth, $\lambda$, in type II superconductors. The magnetic penetration depth is one of the most fundamental parameters in a superconductor since it is related to the superfluid density, $n_s$, via 1/${\lambda}^{2}$=${\mu}_{0}$$e^{2}$$n_{\rm s}$/$m^{*}$ (where $m^*$ is the effective mass). In addition, the temperature dependence of $1/\lambda^2$ is related to the pairing symmetry of the superconductor, and therefore can provide information on the presence of multiband superconductivity or nodes. 

Firstly, TF-$\mu$SR measurements on Sc$_6$FeTe$_2$ (Fig. 2a) above (black, $6~$K) and below (red, $0.03~$K) $T_\mathrm{c}$ show the expected response from a superconductor; a weakly damped oscillation above $T_\mathrm{c}$ due to the random local fields produced by the nuclear moments, which is strongly enhanced below $T_\mathrm{c}$ due to the formation of the flux line lattice (FLL). This is further apparent in Fig. \ref{Fig2}d, which shows the Fourier transform (FT) of these $\mu$SR spectra, which is sharp and symmetrical above $T_\mathrm{c}$, but broadens below $T_\mathrm{c}$ into an asymmetrical lineshape that was analysed with the following functional form for $i$ components \cite{musrfit,maisuradze2009comparison}:
\begin{equation}
    A_\mathrm{TF}(t) = \sum_i^n A_{S,i}e^ {\left[ - \frac{(\sigma_{\mathrm{sc},i}+\sigma_{\mathrm{nm}})t^2}{2} \right]}\mathrm{cos}(\gamma_\mu B_{\mathrm{int},i}t + \phi_i)
\end{equation}
where $A_\mathrm{S}$ is the initial asymmetry, $\sigma_\mathrm{sc}$ and $\sigma_\mathrm{nm}$ are muon spin depolarisation rates, $\gamma_\mu/(2\pi) \simeq 135.5~$MHz/T is the gyromagnetic ratio of the muon, $B_\mathrm{int}$ is the internal magnetic field, and $\phi$ is the initial phase shift of the muon ensemble. The relaxation rates $\sigma_\mathrm{sc}$ and $\sigma_\mathrm{nm}$ characterise the damping due to the formation of the FLL in the superconducting state and the nuclear magnetic dipolar contribution, respectively. Since above $T_\mathrm{c}$ there is no superconducting contribution remaining, $\sigma_\mathrm{nm}$ is obtained by averaging rates above $T_\mathrm{c}$ where only nuclear moments contribute to the muon depolarisation rate. The remaining $\sigma_\mathrm{sc}$ is a direct measure of the penetration depth and, consequently, the superfluid density (see Eq. 2 of the extended data section).

The responses in the superconducting state of Sc$_6$RuTe$_2$ and Sc$_6$IrTe$_2$ are a bit different. Namely, only a very weak damping is observed below $T_\mathrm{c}$ (Figs. \ref{Fig2}b and c, respectively), indicating that the superconducting relaxation rate is very small. This minimal increase in damping could be attributed to two potential factors: (1) the superconducting volume fraction in these samples is small, or (2) the penetration depth is significantly larger in these systems. The standard method for extracting $\sigma_\mathrm{sc}$ from $\mu$SR spectra in type-II superconductors involves field-cooling the sample in a magnetic field to create a well-ordered FLL. However, it is well-known that disorder in the vortex lattice increases the distribution of internal magnetic fields, thereby enhancing the damping. Deliberately introducing disorder into the vortex lattice thus provides a useful approach to probe the superconducting volume fraction of the crystals. A widely established procedure to induce such disorder involves cooling the sample below $T_\mathrm{c}$ in zero field, followed by sweeping the magnetic field to the desired transverse-field, a method referred to as zero-field cooling (ZFC). To distinguish between the effects of a large penetration depth and a reduced superconducting volume fraction, we conducted ZFC measurements on both Ru and Ir samples of Sc$_6M$Te$_2$. Under ZFC conditions, a strong damping of the full signal was observed in both cases proving the bulk character of the superconductivity. Based on these results, we conclude that the small relaxation in the superconducting states of the Ru and Ir samples arises from the much larger penetration depths in these systems compared to the Fe sample. The Ru and Ir Sc$_6M$Te$_2$ data were also analysed using Eq. 1, with a single component for the field cooled (FC) data, two components for the ZFC Ru sample and four components for the ZFC Ir sample. The small peak near $0~$mT in Fig. \ref{Fig2}f can be attributed to the expulsion of the magnetic field from some sample regions due to the Meissner effect \cite{amato2024introduction}. All subsequent analyses will be performed on the FC data, as this mode enables the extraction of damping from a well-ordered vortex lattice, providing a reliable determination of $\lambda^{-2}$.

\begin{figure*}
    \centering
    \includegraphics[width=\textwidth]{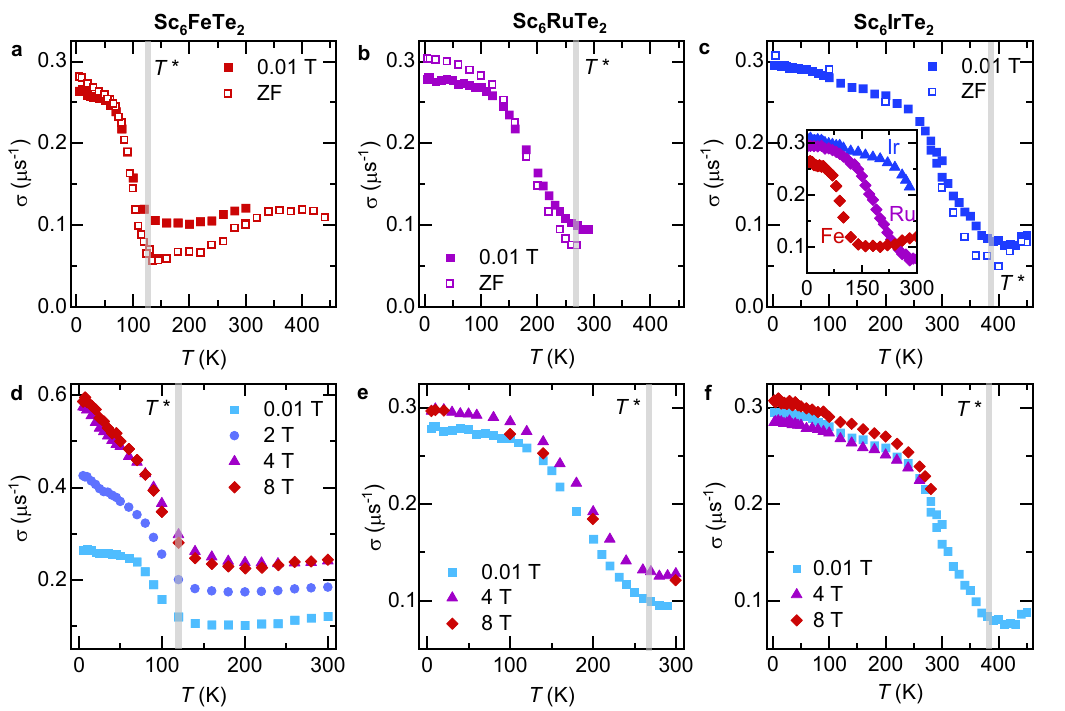}
    \caption{\textbf{Summary of $\mu$SR experiments in the normal state of Sc$_6M$Te$_2$ ($M =$ Fe, Ru and Ir)  a - c} Zero-field (ZF, open markers) and TF (closed markers)-$\mu$SR measurements as a function of temperature for Sc$_6$FeTe$_2$, Sc$_6$RuTe$_2$ and Sc$_6$IrTe$_2$, respectively. The strong upturn of the relaxation rate is denoted as the $T^*$ transition, which varies significantly with $M$ substitution (\textbf{c}. inset). \textbf{d - f} Measurements of the relaxation rate under applied fields of 0.01 to $8~$T for Sc$_6$FeTe$_2$, Sc$_6$RuTe$_2$ and Sc$_6$IrTe$_2$, respectively.}
    \label{fig4}
\end{figure*}

The temperature dependence of $\sigma_\mathrm{sc}$ reflects the topology of the superconducting pairing symmetry, therefore we conducted a quantitative comparison of our data with the most common superconducting gap structures, where a full analysis and discussion can be found in the extended data. The dominant plateau in the Sc$_6$FeTe$_2$ data below $1~$K would suggest that a nodeless $s$-wave pairing symmetry (Figure \ref{fig3}a solid line) would be the most appropriate, however, it is clear that this is not the correct model. An extension to the simple $s$-wave is to consider either two constant gaps in the Fermi surface ($s+s$) or an anisotropic $s$-wave which has a radial dependence but does not at any point go to zero and so remains nodeless. These models are shown by the dashed black and grey lines in Fig. \ref{fig3}a and are indistinguishable.
\begin{table}
    \centering
    \begin{tabular}{c|ccc} 
    \hline
         Sample&  Fe&  Ru& Ir\\ 
         \hline
         $\lambda^{-2} (T = 0) (\mu$m$^{-2}$)&  7.78(5)&  1.22(3)& 0.83(2)\\ 
         $\lambda (T > 0)$ (nm)&  360&  905& 1100\\ 
         $\omega$&  0.68(5)&  1& 1\\ 
         $T_\mathrm{c}$ (K)&  4.00(2)&  1.9(1)& 1.6(1)\\ 
         $\Delta_1$ (meV)&  1.3(1)&  0.44(6)& 0.33(4)\\ 
         $\Delta_2$ (meV)&  0.35(4)&  -& -\\ 
         \hline
    \end{tabular}
    \caption{\textbf{Summary of superconducting gap structure parameters for the Fe, Ru and Ir Sc$_6M$Te$_2$ samples.} The Fe sample was fit with a double $s+s$-wave model, whereas the Ru and Ir samples were fit with a single $s$-wave model. $\lambda$ is the London penetration depth, $\omega$ is the phase fraction, and $\Delta$ is the size of the gap.}
    \label{tab1}
\end{table}
A similar shape to the superfluid density can be found for both Sc$_6$RuTe$_2$ and Sc$_6$IrTe$_2$, but here the simplest $s$-wave gap structure fits well in both cases. Subsequently, we assume that in Sc$_6$FeTe$_2$ the most appropriate description of the gap structure is a double nodeless $s+s$-wave gap, and that as the electron correlations diminish from $3d$ to $5d$ systems, one of these gaps closes. Furthermore, first principle calculations show that in addition to the overlap for the scandium orbitals, in $3d$ transition metals there is a significant contribution to the electronic states arising from the $3d$ orbitals \cite{Sc6MTe2}. It follows that this additional contribution may manifest as the second gap. A summary of the main parameters extracted from the fits are shown in Table \ref{tab1}. The most intriguing result is the evolution of the London penetration depth, $\lambda$, which increases from a modest $360~$nm for the $3d$ Sc$_6$FeTe$_2$ up to an extraordinarily large $1100~$nm for the $5d$ Sc$_6$IrTe$_2$. These results quantitatively support the conclusion that the small damping of the asymmetry signal in Fig. \ref{Fig2}b and c was due to a long $\lambda$. More importantly, the long penetration depth $\lambda$---indicative of a dilute superfluid density---strongly suggests that the Sc$_6M$Te$_2$ family belongs to the class of unconventional superconductors. This is further supported by the observed correlation between superfluid density and $T_\mathrm{c}$, a well-established experimental hallmark of unconventional superconductivity.

Finally, Fig. \ref{fig3}b shows the temperature dependence of the internal magnetic field, $\mu_0H_\mathrm{int}$, for each of the systems, which display contrasting behaviours. Firstly, Sc$_6$FeTe$_2$ experiences a large diamagnetic shift of $H_\mathrm{int}$ which is to be expected in the superconducting state. The response from Sc$_6$RuTe$_2$ and Sc$_6$IrTe$_2$ are significantly weaker, and corresponds to a very weak diamagnetic shift for Sc$_6$RuTe$_2$ and an unusual paramagnetic shift for Sc$_6$IrTe$_2$.  This paramagnetic shift may be caused by field induced magnetism \cite{PhysRevLett.103.067010, PhysRevB.93.094513}, vortex disorder \cite{PhysRevLett.106.127002}, demagnetisation, an odd-superconducting pairing \cite{PhysRevLett.125.026802} or the suppression of the negative Knight shift below $T_\mathrm{c}$ due to singlet pairing. Alternatively, since the experiments were conducted on polycrystalline samples, the very weak response of the Sc$_6$RuTe$_2$ and Sc$_6$IrTe$_2$ internal fields may be due to some inhomogeneity between different grains of the sample where some regions may be paramagnetic which counteracts the response of the internal magnetic field \cite{PhysRevB.72.104504}.

\section{Normal State}
Unconventional superconductivity is often accompanied by a competing or co-operative phase in the normal state that may exist orders of magnitudes above $T_\mathrm{c}$. Through our investigations into the Sc$_6M$Te$_2$ family, we have found new evidence to suggest that such a phase exists in each of the $M =$ Fe, Ru and Ir compounds.

Firstly, Fig. \ref{fig4}a - c shows a comparison between zero-field (ZF) and weak TF ($0.01~$T) $\mu$SR measurements for Sc$_6$FeTe$_2$, Sc$_6$RuTe$_2$ and Sc$_6$IrTe$_2$, respectively. The zero-field $\mu$SR signal exhibits no precession, instead, it is characterised by weak damping, which shows a significant change with the onset temperature ranging from $T^* = 120~$K for Sc$_6$FeTe$_2$ to $T^* = 260~$K for Sc$_6$RuTe$_2$ and  $T^* = 390~$K for Sc$_6$IrTe$_2$. It is notable that the transition temperature, $T^*$ is directly anticorrelated to $T_\mathrm{c}$, and therefore seemingly controlled by the varying strength of spin-orbit coupling in the $d$-electron metal. Although these results indicate the formation of a secondary phase, from these ZF and TF $\mu$SR measurements alone, we are not able to identify if it arises from a magnetic, electronic or structural origin.

\begin{figure}
    \centering   
    \includegraphics[]{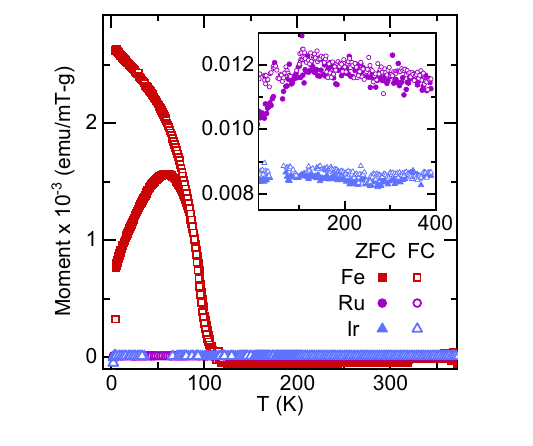}
    \caption{\textbf{Magnetic susceptibility data for the Sc$_6M$Te$_2$ ($M =$ Fe, Ru and Ir) compounds.} Magnetic susceptibility data were collected under zero-field cooled (ZFC) and field cooled (FC) conditions in an applied field of $5~$mT. The Sc$_6$FeTe$_2$ sample shows a clear response below $T^* = 120~$K, but the Sc$_6$RuTe$_2$ and Sc$_6$IrTe$_2$ samples do not have any significant features.}
    \label{fig5}
\end{figure}

To determine the origin of the transition we have conducted additional TF $\mu$SR, neutron diffraction and magnetic susceptibility measurements on the Sc$_6M$Te$_2$ compounds. For the $3d$ Sc$_6$FeTe$_2$ sample, we found a strong increase in the muon spin relaxation rate under applied magnetic fields which saturates at $4~$T (Fig. \ref{fig4}d). The evolution with field does not appear to change the onset of $T^* ( = 120~$K), but does alter the form as the shape becomes much more linear at low temperatures as the field is increased. Magnetic susceptibility measurements (Fig. \ref{fig5}) also show a strong increase below $T^*$ and a clear splitting between the FC and ZFC data. These results therefore indicate that the transition in Sc$_6$FeTe$_2$ is magnetic in nature. However, neutron diffraction measurements (see Extended Data) revealed no additional magnetic Bragg peaks or intensity, suggesting that the magnetism is characterised by extremely weak moments. This likely explains the only slight increase in the muon spin relaxation rate and the absence of spontaneous oscillations. Conversely, the muon spin relaxation rate of the $4d$ Sc$_6$RuTe$_2$ and $5d$ Sc$_6$IrTe$_2$ samples are essentially field independent (Figs. \ref{fig4}e and f, respectively), and have no response in the magnetic susceptibility measurements, which excludes a magnetic transition. Furthermore, the neutron diffraction study could find no evidence for a structural phase transition within the resolution of the instrument (further details in the extended data). It is possible that the transition is electronic in origin, however further measurements by a charge susceptible probe, such as ARPES or RIXS, are necessary.

\section{Summary}
\begin{figure}
    \centering
\includegraphics[]{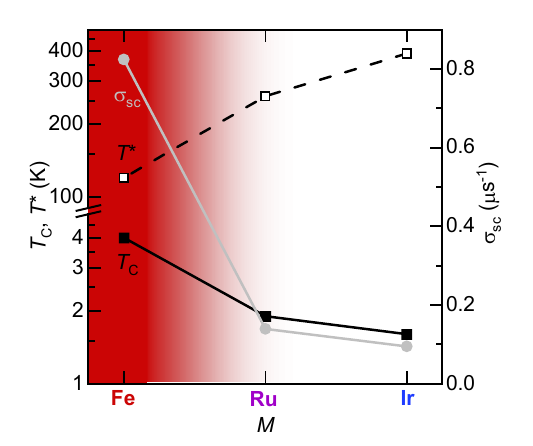}
    \caption{\textbf{Summary of transition temperatures for the Sc$_6M$Te$_2$ compounds.} Critical ($T_\mathrm{c}$, closed) and normal state ($T^*$, open) transitions for the Fe, Ru and Ir Sc$_6M$Te$_2$ compounds. Temperature is shown on a log scale. Additionally, the superfluid density $\sigma_\mathrm{sc} (T = 0)$ is shown on the right axis (light grey circle markers) which has the same evolution as $T_\mathrm{c}$. The red background shows the transition from two to one nodeless gaps as electron correlations decrease.}
    \label{fig6}
\end{figure}

Taken together, these results show that the Sc$_6M$Te$_2$ family provide an experimentally achievable route to tuning superconductivity through the substitution of $d$-electron metals. The fact that the chemical substitution does not change the crystal symmetry is a unique aspect to the Sc$_6M$Te$_2$ family, offering for the first time the chance to perform a truly systematic study of the evolution of $d$-electron superconductivity across $3d$, $4d$ and $5d$ electron systems. Our findings underscore that the unconventional properties of both the superconducting and normal states can be profoundly influenced by the substitution of different $d$-electron metals, reflecting the interplay between the strength of electron correlations and spin-orbit couplings. This includes the superfluid density, which remains dilute in all cases. Notably, the superfluid density correlates with the superconducting critical temperature, exhibiting a significantly stronger correlation in the $3d$ Fe compound with pronounced electron correlations. In contrast, the $5d$ Ir compound displays an extraordinarily long London penetration depth ($1100~$nm) and an unusual paramagnetic shift which may be associated with its strong spin-orbit coupling. The superconducting gap symmetry was best described by a double $s$-wave gap for the $3d$ Fe system, which transitions to a single nodeless gap as electron correlations decrease. Additionally, the two dominant temperature scales, $T_\mathrm{c}$ and $T^*$ (normal state transition) both have a linear dependence but are directly anticorrelated (Fig. \ref{fig6}) which implies that electron correlations co-operate with superconducting strength whereas spin-orbit coupling competes. These findings have significant implications for future materials design, suggesting that while both strong electron correlations and spin-orbit coupling can give rise to unconventional features, maximising electron correlations whilst minimising spin-orbit coupling could be important for achieving the highest superconducting critical temperatures. The high temperature $T^*$ transition in the $3d$ Sc$_6$FeTe$_2$ compound has a magnetic origin, characterised by a small magnetic moment. In contrast, this transition is neither structural nor magnetic in the $4d$ Sc$_6$RuTe$_2$ or $5d$ Sc$_6$IrTe$_2$ systems. The possibility of charge order in these compounds warrants further investigation, with techniques such as ARPES, high-intensity X-ray diffraction or RIXS being suitable methods. 


Finally, the maximum superconducting critical temperature that could be achieved through chemical substitution is $4.5~$K, however a crucial question to ask is whether $T_\mathrm{c}$ can be increased further. Given that the normal state $T^*$ transition is anticorrelated to $T_\mathrm{c}$, one approach could be to suppress $T^*$ using hydrostatic pressure or strain which may then increase $T_\mathrm{c}$ as has been the case in other $d$-electron superconductors such as the $A$V$_3$Sb$_5$ compounds \cite{AV3Sb5_pressure_1, AV3Sb5_pressure_3} or the cuprates \cite{guguchia2020using, guguchia2024designing}. 

\bibliography{References}{}

\section{Acknowledgments}~
Z.G. acknowledges support from the Swiss National Science Foundation (SNSF) through SNSF Starting Grant (No. TMSGI2${\_}$211750). M.M. would like to thank the Swiss National Science Foundation (Grant No. $206021\textunderscore139082$) for funding of the MPMS. Y.O. and K.Y. acknowledge support from the Japan Science and Technology Agency through JST-ASPIRE (No. JPMJAP2314).\\

\section{Author contributions}~
Z.G. conceived and supervised the project. Sample growth: K.Y. and Y.O..
$\mu$SR experiments, the corresponding analysis and discussions: J.N.G., V.S., A.D., C.M.III, P.K., O.G., S.S.I., H.L., and Z.G.. Neutron diffraction and magnetisation experiments: J.N.G., V.P., C. M. III, M.M. and Z.G.. Figure development and writing of the paper: J.N.G. and Z.G. All authors discussed the results, interpretation, and conclusion.\\ 

\section*{Extended Data}
\subsection*{Superconducting gap fits}
To perform a quantitative analysis of $\mu$SR data and determine the superconducting gap structure, the superconducting muon spin depolarisation rate, $\sigma_\mathrm{sc}(T)$ in the presence of a perfect triangular vortex lattice is first related to the London penetration depth, $\lambda(T)$, by the following equation \cite{London_muSR, London_muSR2}:
\begin{equation}
    \frac{\sigma_\mathrm{sc}(T)}{\gamma_\mu} = 0.06091 \frac{\Phi_0}{\lambda^2(T)}
\end{equation}
where $\Phi_0 = 2.068 \times 10^{15}~$Wb is the magnetic flux quantum. This equation is only applicable when the separation between vortices is larger than $\lambda$. In this particular case, as per the London model, $\sigma_\mathrm{sc}$ becomes field-independent. By analysing the temperature of the magnetic penetration depth, within the local London approximation, a direct association with the superconducting gap symmetry can be made \cite{musrfit}:
\begin{equation}
    \frac{\lambda^{-2}(T, \Delta_{0,i})}{\lambda^{-2}(0, \Delta_{0,i})} = 1 + \frac{1}{\pi} \int_0^{2\pi} \int_{\Delta(T, \phi)}^\infty \left(\frac{\delta f}{\delta E} \right) \frac{EdEd\varphi}{\sqrt{E^2-\Delta_i(T,\varphi)^2}}
\end{equation}
where $f = [1+\mathrm{exp}(E/k_BT)]^{-1}$ is the Fermi function, $\varphi$ is the angle along the Fermi surface, and $\Delta_i(T, \varphi) = \Delta_{0,i}\Gamma(T/T_\mathrm{C})g(\varphi)$ ($\Delta_{0,i}$ is the maximum gap value at $T = 0$). The temperature dependence of the gap is approximated by the expression, $\Gamma(T/T_\mathrm{C} = \mathrm{tanh}{1.82[1.018(T_\mathrm{C}/T - 1)]^{0.51}}$ \cite{sc_gap}, whilst $g(\varphi)$ describes the angular dependence of the gap and is replaced by $1$ for an $s$-wave gap, $[1+a\mathrm{cos}(4\varphi)/(1 + a)]$ for an anisotropic $s$-wave gap, and $|\mathrm{cos}(2\varphi)|$ for a $d$-wave gap \cite{sc_gap2}. An $s+s$-wave gap involves two singular $s$-wave gaps that share the same $T_\mathrm{c}$.\\ 
\begin{figure*}
    \centering
    \includegraphics[width=\textwidth]{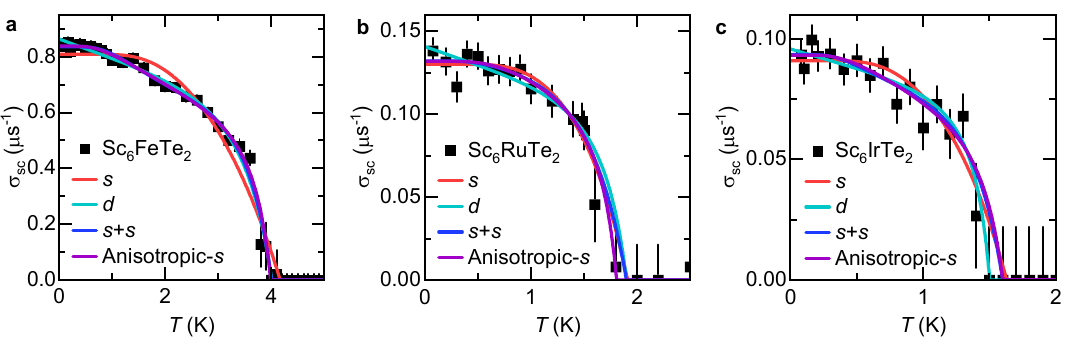}
    \caption{\textbf{Comparison of different superconducting gap structures for the Fe, Ru and Ir Sc$_6M$Te$_2$ samples. }Models included are a single $s$-wave (red), $d$-wave (light blue), $s+s$-wave (dark blue) and anisotropic-$s$-wave (purple).}
    \label{Ext_fig1}
\end{figure*}
\subsection*{Sc$_6$FeTe$_2$}
Fits of the superconducting gap structure for Sc$_6$FeTe$_2$ are summarised in Fig. \ref{Ext_fig1}a and Table \ref{Ext_tab_1}. Firstly, the single $s$-wave model was discounted because the shape does not fit the experimental data well. Next, the nodal $d$-wave model was discounted as the data have a dominant plateau below $1~$K, which is suggestive of a nodeless superconducting gap structure. The expected temperature dependence of $\lambda^{-2}$ for a nodal $d$-wave structure is a continually almost linear increase down to $0~$K. That leaves the double nodeless $s+s$ and anisotropic $s$ models which as can be seen in Fig. \ref{Ext_fig1}a are indistinguishable from each other. The reason why the $s+s$-wave model was selected in the main text is due to the more natural explanation that moving from $3d$ to $4d$ to $5d$ systems, does not change the fundamental nature of the gap but is due to one gap simply closing as the electron correlations diminish. This will have to be confirmed with other techniques, such as ARPES.
\begin{table}
    \centering
    \begin{tabular}{c|cccc} 
    \hline
         Model&  $s$&  $d$& $s+s$&Anisotropic-$s$\\ 
         \hline
         $\lambda^{-2} (T = 0) (\mu$m$^{-2}$)&  7.53(3)&  8.04(4)& 7.78(5)
&7.81(5)\\ 
         $\lambda (T > 0)$ (nm)&  365&  355& 360
&360\\ 
         $\omega$&  1&  1& 0.68(5)
&1\\ 
         $T_\mathrm{c}$ (K)&  4.17(4)&  4.01(2)& 4.00(2)
&3.99(6)\\ 
         $\Delta_1$ (meV)&  0.78(2)&  1.48(3)& 1.3(1)
&1.11(6)\\ 
         $\Delta_2$ (meV)&  -&  -& 0.35(4)
&-\\
 $a$& -& -& -&0.78(3)\\ 
&26.91\\ 
         \hline
    \end{tabular}
    \caption{\textbf{Summary of different superconducting gap structure models for Sc$_6$FeTe$_2$}.}
    \label{Ext_tab_1}
\end{table}
\subsection*{Sc$_6$RuTe$_2$}
Fits of the superconducting gap structure for Sc$_6$RuTe$_2$ are summarised in Fig. \ref{Ext_fig1}b and Table \ref{Ext_tab_2}. Similarly to Sc$_6$FeTe$_2$, the data plateau below $1~$K which excludes the nodal $d$-wave model. The other models---single $s$, double $s+s$ and anisotropic $s$---are all quantitatively similar, both in terms of the London penetration depth and gap size. The double $s+s$ however, in this case, can be excluded as the error on the second gap ($\Delta_2$) and phase fraction ($\omega$) are outside the confidence level. This leaves $s$ and anisotropic $s$, which as we have described in the main text are indistinguishable but we have chosen $s$ as the most likely model by assuming that the larger gap in Sc$_6$FeTe$_2$ has closed.
\begin{table}[h!]
    \centering
    \begin{tabular}{c|c c c c} 
    \hline
         Model&  $s$&  $d$& $s+s$& Anisotropic-$s$\\ 
         \hline
         $\lambda^{-2} (T = 0) (\mu$m$^{-2}$)&  1.22(3)&  1.32(3)& 1.24(4)&1.24(4)\\ 
         $\lambda (T > 0)$ (nm)&  905&  870& 900&900\\ 
         $\omega$&  1&  1& 0.9(7)&1\\ 
         $T_\mathrm{c}$ (K)&  1.9(1)&  1.9 (fixed)& 1.9 (fixed)&1.81(5)\\ 
         $\Delta_1$ (meV)&  0.44(6)&  0.75(9)& 0.5(2)&0.5(1)\\ 
         $\Delta_2$ (meV)&  -&  -& 0.2(2)&-\\
 $a$& -& -& -&0.6(2)\\ 
         \hline
    \end{tabular}
    \caption{\textbf{Summary of different superconducting gap structure models for Sc$_6$RuTe$_2$}.}
    \label{Ext_tab_2}
\end{table}

\subsection*{Sc$_6$IrTe$_2$}
Fits of the superconducting gap structure for Sc$_6$IrTe$_2$ are summarised in Fig. \ref{Ext_fig1}c and Table \ref{Ext_tab_3}. Similarly to the Sc$_6$FeTe$_2$ and Sc$_6$RuTe$_2$ samples, all fits in Fig. \ref{Ext_fig1}c are very close together and the models are difficult to differentiate between. Sc$_6$IrTe$_2$ is also the sample with the most dilute superfluid density which makes the errors comparatively the largest. The data plateau below $0.5~$K and so the $d$-wave model was discounted. This leaves the three nodeless models, and like for Sc$_6$RuTe$_2$ we have assumed that the simplest single $s$-wave structure is the correct model. This model suggests a slight suppression of the gap from the Sc$_6$RuTe$_2$ sample, which would be expected as the spin-orbit coupling is increased.

\begin{table}[h!]
    \centering
    \begin{tabular}{c|c c c c} 
    \hline
         Model&  $s$&  $d$& $s+s$& Anisotropic-$s$\\ 
         \hline
         $\lambda^{-2} (T = 0) (\mu$m$^{-2}$)&  0.83(2)&  0.87(2)& 0.85(3)&0.85(3)\\ 
         $\lambda (T > 0)$ (nm)&  1100&  1070& 1085&1085\\ 
         $\omega$&  1&  1& 0.7(2)&1\\ 
         $T_\mathrm{c}$ (K)&  1.6(1)&  1.50(4)& 1.6 (fixed)&1.6 (fixed)\\ 
         $\Delta_1$ (meV)&  0.33(4)&  0.7(1)& 0.5(2)&0.41(8)\\ 
         $\Delta_2$ (meV)&  -&  -& 0.1(1)&-\\
 $a$& -& -& -&0.7(2)\\ 
         \hline
    \end{tabular}
    \caption{\textbf{Summary of different superconducting gap structure models for Sc$_6$IrTe$_2$}.}
    \label{Ext_tab_3}
\end{table}
\subsection*{Neutron diffraction}
\begin{figure}
    \centering
    \includegraphics[width=\linewidth]{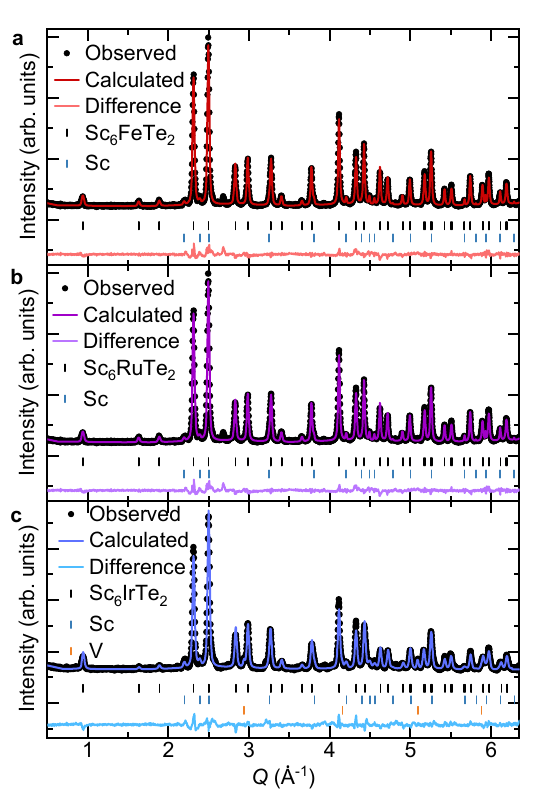}
\caption{\textbf{Example Rietveld refinement of neutron data from HRPT for the Fe, Ru and Ir Sc$_6M$Te$_2$ samples.} All data were collected at $1.5~$K with a wavelength, $\lambda = 1.89~$\r{A}.}
    \label{Ext_fig2}
\end{figure}
Neutron diffraction data collected on the Fe, Ru and Ir Sc$_6M$Te$_2$ samples on the HRPT instrument\cite{fischer2000high} at the Paul Scherrer Institute are summarised in Fig. \ref{Ext_fig2}. Data were collected between $1.5$ and $300~$K for Sc$_6$FeTe$_2$ and Sc$_6$RuTe$_2$, and $1.5$ and $450~$K for the Sc$_6$IrTe$_2$ sample. All data were analysed using the Fullprof package. A combined refinement of $\lambda = 1.89~$\r{A} and $\lambda = 1.15~$\r{A} was completed at $1.5~$K for each sample, and then all the instrument parameters (zero, peak widths, absorption corrections, radial dependences) were kept constant for the rest of the refinement. The only parameters refined as a function of temperature were the lattice parameters, isotropic thermal parameters, $B_\mathrm{iso}$, and background. All samples contained a small $\sim 4~\%$ Sc impurity (blue tickmarks) and an unidentified trace impurity. Vanadium peaks from the sample can were modelled with the le Bail method (orange tickmarks) for the Sc$_6$IrTe$_2$ sample.

\begin{figure}
    \centering
    \includegraphics[width=\linewidth]{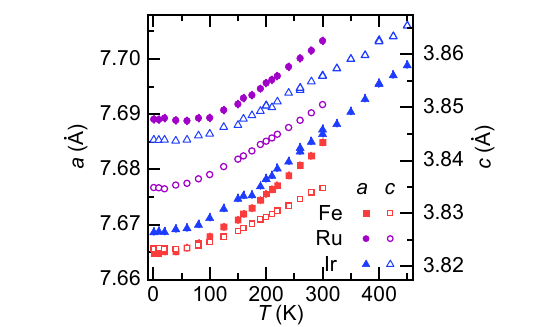}
    \caption{\textbf{Summary of lattice parameters for the Fe, Ru and Ir Sc$_6M$Te$_2$ samples.} All samples show the expected lattice compression and there are no anomalies to suggest a structural phase transition. Solid and open markers describe the $a (=b)$ and $c$ lattice parameters on the left and right axes, respectively.}
    \label{Ext_fig3}
\end{figure}

An example fit for each of the Sc$_6M$Te$_2$ samples is shown in Fig. \ref{Ext_fig2} with data collected at $1.5~$K. In all cases the Sc$_6M$Te$_2$ was well modelled by the previously reported hexagonal $P\bar{6}2m$ structure at all temperatures \cite{maggard2000sc6mte2, Sc6MTe2}. A summary of the lattice parameters can be found in Fig. \ref{Ext_fig3}, which show the expected compression of $a$ and $c$ with thermal contraction. A summary of structural parameters can be found in Tables \ref{Ext_tab_4}, \ref{Ext_tab_5} and \ref{Ext_tab_6} for the Fe, Ru and Ir Sc$_6M$Te$_2$ samples, respectively. We found no evidence for significant local distortions to the Sc$_6M$ octahedra. The Sc$_6$IrTe$_2$ sample has slightly lower goodness of fit parameters due to the increased background from the incoherent scattering of iridium. We found no evidence for additional magnetic Bragg peaks/intensity at low temperatures.
\begin{table}
    \centering
    \begin{tabular}{cccccc}
    \hline
         Atom&  Site&  $x$&  $y$&  $z$& B(\r{A}$^{-2}$)\\
         \hline
         Sc1&  $3g$&  0.23677(2)&  0&  0.5& 0.48(1)\\
         Sc2&  $3f$&  0.61468(2)&  0&  0& 0.48(1)\\
         Fe&  $1a$&  0&  0&  0& 0.35(4)\\
         Te&  $2d$&  0.3333&  0.6667&  0& 0.10(3)\\
         \hline
    \end{tabular}
    \caption{\textbf{Refined structural parameters for Sc$_6$FeTe$_2$ from HRPT data collected at $1.5~$K.} Structure was refined using the hexagonal $P\bar{6}2m$ spacegroup, with refined lattice parameters, $a = b = 7.66483(3)~$\r{A} and $c = 3.82333(3)~$\r{A}. Goodness of fit parameters are $R_\mathrm{wp} = 7.78~\%$ and $\chi^2 = 3.76$.}
    \label{Ext_tab_4}
\end{table}
\begin{table}
    \centering
    \begin{tabular}{cccccc}
    \hline
         Atom&  Site&  $x$&  $y$&  $z$& B(\r{A}$^{-2}$)\\
         \hline
         Sc1&  $3g$&  0.23690(2)&  0&  0.5& 0.77(1)\\
         Sc2&  $3f$&  0.61170(3)&  0&  0& 0.77(1)\\
         Ru&  $1a$&  0&  0&  0& 1.17(4)\\
         Te&  $2d$&  0.3333&  0.6667&  0& 0.17(2)\\
         \hline
    \end{tabular}
    \caption{\textbf{Refined structural parameters for Sc$_6$RuTe$_2$ from HRPT data collected at $1.5~$K.} Structure was refined using the hexagonal $P\bar{6}2m$ spacegroup, with refined lattice parameters, $a = b = 7.68904(3)~$\r{A} and $c = 3.83486(3)~$\r{A}. Goodness of fit parameters are $R_\mathrm{wp} = 11.8~\%$ and $\chi^2 = 3.44$.}
    \label{Ext_tab_5}
\end{table}
\begin{table}
    \centering
    \begin{tabular}{cccccc}
    \hline
         Atom&  Site&  $x$&  $y$&  $z$& B(\r{A}$^{-2}$)\\
         \hline
         Sc1&  $3g$&  0.24128(3)&  0&  0.5& 1.28(3)\\
         Sc2&  $3f$&  0.61017(1)&  0&  0& 1.28(3)\\
         Ir&  $1a$&  0&  0&  0& 3.1(1)\\
         Te&  $2d$&  0.3333&  0.6667&  0& 0.10(5)\\
         \hline
    \end{tabular}
    \caption{\textbf{Refined structural parameters for Sc$_6$IrTe$_2$ from HRPT data collected at $1.5~$K.} Structure was refined using the hexagonal $P\bar{6}2m$ spacegroup, with refined lattice parameters, $a = b = 7.66867(3)~$\r{A} and $c = 3.84362 (3)~$\r{A}. Goodness of fit parameters are $R_\mathrm{wp} = 13.9~\%$ and $\chi^2 = 5.56$.}
    \label{Ext_tab_6}
\end{table}

Therefore, these results confirm there are no structural or long-range ordered magnetic transitions within the resolution of the instrument. We are not able to rule out an electronic origin to $T^*$, however we would require X-ray diffraction measurements, preferably on single-crystals, in order to see small changes to the structure that may arise from charge order. 

\subsection*{Longitudinal-field and zero-field $\mu$SR}
\begin{figure*}
    \centering
    \includegraphics[width=\textwidth]{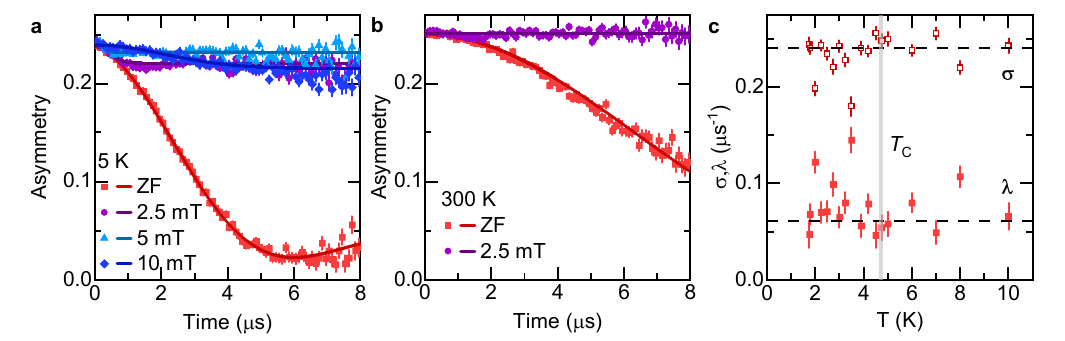}
    \caption{\textbf{Summary of Longitudinal-field (LF) and zero-field (ZF) $\mu$SR measurements of Sc$_6$FeTe$_2$ in the normal state.} ZF and LF $\mu$SR data collected at \textbf{a} $5~$K and \textbf{b} $300~$K in various applied fields. \textbf{c} ZF measurements across $T_\mathrm{c}$ show no increase which suggests the absence of time-reversal symmetry breaking.}
    \label{Ext_fig4}
\end{figure*}
Longitudinal-field (LF) and zero-field (ZF) $\mu$SR measurements are shown in Fig. \ref{Ext_fig4} for the Sc$_6$FeTe$_2$ sample only. At $5~$K, the ZF spectra in Fig. \ref{Ext_fig4}a has a standard Gaussian Kubo-Toyade function, convoluted with a exponential term \cite{musrfit}:
\begin{equation}
    P_\mathrm{ZF}^\mathrm{GKT}(t) = \left( \frac{1}{3} + \frac{2}{3}(1-\sigma_i^2t^2)\mathrm{exp}\left[- \frac{\sigma_i^2t^2}{2}\right]\right) \mathrm{exp}(-\Gamma_i t)
\end{equation}
where $\sigma_i$ and $\Gamma_i$ are the muon spin relaxation rates. Following the application of a small LF, the nuclear moments are largely decoupled, indicating that the relaxation is due to spontaneous fields which are static on the microsecond timescale. However a small depolarisation persists, which actually appears to get stronger with the application of the LF. This is unusual but may suggest that there is an electronic component to the relaxation which becomes more prominent with the decoupling of the nuclear moments. This will have to be explored further with other techniques. Additional measurements were also performed at $300~$K (Fig. \ref{Ext_fig4}b) and show a complete decoupling of the nuclear moments with a LF of $2.5~$mT. The extra electronic component does not appear to be present at $300~$K ($> T^*$). Finally, we measured ZF data across $T_\mathrm{c}$, and find no evidence for a weak increase in the rate. This indicates the absence of time-reversal symmetry breaking in the superconducting state.
\end{document}